\documentclass{article}

\usepackage{arxiv}

\usepackage[utf8]{inputenc} 
\usepackage[T1]{fontenc}    
\usepackage{hyperref}       
\usepackage{url}            
\usepackage{booktabs}       
\usepackage{amsfonts}       
\usepackage{nicefrac}       
\usepackage{microtype}      
\usepackage{lipsum}
\usepackage{amssymb}
\usepackage{amsmath}
\usepackage{graphicx}
\usepackage{caption}
\usepackage{multirow}
\usepackage{booktabs}
\usepackage{tabularx}

\title{\textbf{Neural Embeddings of Urban Big Data Reveal Emergent Structures in Cities}}

\author{
    \sffamily\large \vspace{0.15in}
    Chao Fan\textsuperscript{1,*}, Yang Yang\textsuperscript{2}, Ali Mostafavi \textsuperscript{1,*}\\
    \sffamily\normalsize
    \textsuperscript{1}Department of Civil and Environmental Engineering, Texas A\&M University, College Station, TX, 77843, U.S.\\
    \sffamily\normalsize
    \textsuperscript{2}Department of Computer Science and Engineering, Texas A\&M University, College Station, TX, 77843, U.S.\\
    \sffamily\normalsize
    \textsuperscript{*}Corresponding authors: chfan@tamu.edu, amostafavi@civil.tamu.edu\\
}

\begin{document}
\maketitle

\begin{abstract}
\sffamily Over decades, many cities have been expanded and functionally diversified by population activities, socio-demographics and attributes of the built environment. Urban expansion and development have led to emergence of spatial structures of cities. Uncovering cities’ emergent structures and spatial gradients is critical to the understanding of various urban phenomena such as segregation, equality of access, and sustainability. In this study, we propose using a neural embedding model—graph neural network (GNN)— that leverages the heterogeneous features of urban areas and their interactions captured by human mobility network to obtain vector representations of these areas. Using large-scale high-resolution mobility data sets from millions of aggregated and anonymized mobile phone users in 16 metropolitan counties in the United States, we demonstrate that our embeddings encode complex relationships among features related to urban components (such as distribution of facilities) and population attributes and activities. The spatial gradient in each direction from city center to suburbs is measured using clustered representations and the shared characteristics among urban areas in the same cluster. Furthermore, we show that embeddings generated by a model trained on a different county can capture 50\% to 60\% of the emergent spatial structure in another county, allowing us to make cross-county comparisons in a quantitative way. Our GNN-based framework overcomes the limitations of previous methods used for examining spatial structures and is highly scalable. The findings reveal non-linear relationships among urban components and anisotropic spatial gradients in cities. Since the identified spatial structures and gradients capture the combined effects of various mechanisms, such as segregation, disparate facility distribution, and human mobility, the findings could help identify the limitations of the current city structure to inform planning decisions and policies. Also, the model and findings set the stage for a variety of research in urban planning, engineering and social science through integrated understanding of how the complex interactions between urban components and population activities and attributes shape the spatial structures in cities. 
\end{abstract}


\vspace{0.3in}
\newpage
\section*{\sffamily Introduction}
\vspace{-0.5em}
The upshot of population shift and its resultant rapid urbanization occurring over the course of the last few decades is that more than half of the world’s population live in cities \cite{Bassolas2019}. Human activities intertwining with facilities and the environment result in spatial structures and variations in urban areas, such as dispersed hubs of businesses and residents \cite{Zhong2015}, which in turn exert a significant influence on people’s life \cite{DiClemente2018}, social equity \cite{Yang2020}, and sustainable development of cities \cite{Brelsford2018}. For better planning strategies and design of cities, is imperative to understand the spatial structures of cities emerging from the complex interactions among urban components (e.g., facilities and the built environments), population attributes, and activities. 

Understanding the spatial structure of cities has been one of the foundational challenges in urban studies \cite{Li2017}, and a growing and diverse number of studies have been investigating this phenomenon over the past two decades. The most observable phenomenon that researchers first attempt to look into is the form taken by cities. Through the mapping of households \cite{Zhang2016}, economic activities \cite{Zhang2017b}, facilities and infrastructure \cite{Huai2021}, the characteristics of poly-centricity and mono-centricity of cities have been identified in prior studies \cite{Burger2011}. A wide range of analytical approaches have been developed to measure the centrality of the cities. One branch of studies has focused on measuring urban centrality using statistical metrics. For example, Midelfart-Knarvik et al. first proposed a spatial separation index to measure the spatial distribution of industries in urban regions \cite{Midelfart2000}. Built upon that, Pereira et al. introduced the urban centrality index as an extension of the spatial separation index to identify distinct urban structures from different spatial patterns, making the structures between extreme mono-centricity and poly-centricity comparable \cite{Pereira2013}. Centrality metrics are based solely on the distribution of residents and facilities. In more recent studies, the numbers of people attracted to different locations and the diversity of the locations have been considered. Zhong et al. proposed a centrality index to identify functional urban centers, contributing to our understanding of the urban forms and their transformation over time \cite{Zhong2015}. Despite the advancement and usefulness of these metrics, the existing studies do not fully capture the complex interactions among features of urban components (i.e., facilities and built environment) and population activities and attributes giving rise to spatial structures and variations. 

Recognizing this, multiple statistical and mathematical methods, such as network models \cite{Lu2018}, scaling index \cite{Song2010a}, and regression methods \cite{Nilsson2014}, have been employed to analyze the spatial structures of cities \cite{Noulas2012}. The network models are primarily applied to examine the structures of human mobility \cite{Ren2014}, transportation systems \cite{Zhou2007}, and road networks \cite{PhysRevE.96.052301}, which connect areas of cities. Accordingly, researchers employed metrics from network science, such as degree centrality and betweenness centrality, to identify critical areas in the structure \cite{Jia2021}. Moving beyond network configuration and centrality metrics, some previous studies also adopted probability density functions to capture the scaling effects in cities \cite{Lammer2006}. For example, Pan et al. proposed a generative process to simulate the super-linear scaling of social-tie density in cities to uncover the interplay between geography, population density and social interaction \cite{Pan2013}. Yan et al. developed a general model to predict the scaling behaviors of collective mobility in cities \cite{Yan2017a}. In addition, researchers have explored spatial economic models to integrate economic factors in characterizing the spatial structures of cities. These prior studies \cite{Li2014} put more effort into revealing the economic disparities in relation to the geography of urban components, such as high-speed railways \cite{Jin2020}, financial services \cite{Wong2021}, and airports \cite{Chen2021}. While these models allow capture of multiple features of cities and their interactions, they are linear models which can capture only simple linear relationships among features of urban functions and components. Many cities, in fact, have sprawled into metropolitan areas with spatial expansion and functional diversification \cite{Cao2016}. It is therefore not surprising that the nature of the spatial structures of cities is emergent and multifaceted with complex and non-linear relationships between features of urban components and population activities and attributes \cite{Maurer2002,Yan2020}. Because these non-linear relationships influence emergent spatial structure of a city, the forces from the city center to peripheral areas could be anisotropic (i.e., vary non-linearly along different radii from city centers), leading to heterogeneous spatial gradiants \cite{Lin2016a}. Spatial gradient captures the variation of spatial structure across the city, which can further reveal the combined patterns of urban phenomena, such as segregation, disparate facility distribution, and human mobility. Existing studies \cite{Liu2018vertical} on spatial gradients, however, rely primarily on the measurement of the radius from the city center, which could be misleading because spatial structures could vary non-linearly along the same radius. Hence, existing statistical models may not be adequate for capturing multifaceted urban structures and may not interpret gradient patterns properly.

The reasons for the limitations in the existing literature are twofold. First, large-scale data sets to capture diverse features associated with urban components and population activities and their interactions in cities were limited. Hence, prior studies \cite{Cao2016} could draw findings based only on static and course-grained features, such as population size, building types and business investment \cite{Li2019}. Owing to the ubiquity of smart phones and online platforms, a great amount of human-generated and geotagged data capturing the locations, movements, communications, and connections among the populations are collected and shared through privacy-protected agreements. For example, one emerging data source is the locations of mobile phone users who opt-in to share their location information with the data collection companies \cite{Fan2021fine}. These data provide significant value for revealing human mobility behaviors and their interplay with the built environment and geography \cite{Jia2020}, such as the movement flow from one area to another \cite{Ren2014}, visits to a specific type of point of interest \cite{Li2021disparate}, and the colocation probabilities between the residents from two areas \cite{Fan2021colocation}. The availability of such datasets provide the opportunity for encoding highly representative spatial interactions based on human activities in cities. Second, methods that allow characterization of the complex relationships among a variety of urban components and population features are limited. Although interpretable, it is challenging to define statistical and mathematical models without prior knowledge about how the features and their complex interactions shape structures in urban areas. Recent advancements in deep learning have demonstrated that neural embedding techniques offer a powerful solution for non-linear problems \cite{Peng2021}. The idea of embedding approaches is to discern a vector representation for each entity, which can encode the multifaceted relationships between entities \cite{Peng2021}. There has been a large family of embedding approaches, including word embedding \cite{levy2014neural}, document embedding \cite{lau-baldwin-2016-empirical}, item embedding \cite{Barkan2016}, and graph embedding \cite{Scarselli2009}. The most advanced embedding techniques are graph neural networks like node2vec \cite{grover2016} and GraphSage \cite{Hamilton2017}. Different from models that train individual embeddings for each entity, graph neural networks generate embeddings by sampling and aggregating features from a node’s local neighborhood \cite{Hamilton2017}. These algorithms consider not only the entity itself, but also the topological structure of the entire graph \cite{hamilton2017representation}. Since urban areas are composed of spatial areas with different populations and built environment features, both connected by population activities, graph neural networks fit the needs of downstream tasks in urban studies well.

Prior urban studies \cite{Liu2019} involving neural networks achieved two major tasks. One was the prediction of urban phenomena, such as air pollution \cite{Cabaneros2019}, traffic flows \cite{Peng2020} and floods \cite{yuan2021spatiotemporal}. For example, Chan et al. proposed a neural network-based knowledge discovery system to estimate the air pollution levels in urban areas using the spatial-temporal information of air-related factors \cite{YanChan2013}. Another major capability of neural networks is classification and labeling of urban areas based on their functionalities, including land use \cite{Weng2017} and street classification \cite{Wang_Li_Rajagopal_2020}. Recently, Hu et al. classified urban functions at the road-segment level using taxi trajectory data and a graph convolutional neural network \cite{Hu2021}. These prior studies demonstrated the performance of novel deep learning models in capturing the non-linear relationships among urban components with diverse large-scale data sets. Despite the successful applications of graph neural networks in these two major downstream tasks in urban studies, little is known about the multifaceted structures and anisotropic gradients of cities based on the consideration of non-linear interactions of urban components and their associated features.

In this study, we propose and implement a framework to obtain compact representations of urban areas by considering multiple features related to urban components and population attributes and activities to uncover the characteristics of spatial structures and gradients. With a grid overlaying on a county map, we first extract features, including socio-demographic contexts, facility services, population movements, and building types for each grid cell from US census survey data, anonymized mobile phone data, and points of interest data. We construct an adjacency matrix and a feature matrix for each grid cell as the inputs of the neural network model. Then, a graph neural network and a DistMult layer are used, assembled and calibrated to learn the embeddings of grid cells. To validate the embeddings generated by the model, we tested the performance of the model on a classification task in which the model is used to predict the level of movement flow between two grid cells. The embeddings are included in further analyses only when the predictive performance of the model is acceptable. Using the neural embeddings of the urban areas, we cluster these areas based on the distances between the embedding vectors. The spatial structures of cities can be characterized through an understanding of the features in each cluster of urban areas. The embeddings capture the nature of the spatial expansion and multifaceted structures of cities. Our results also demonstrate anisotropic gradients in different directions from the city center to the peripheral areas. We further hypothesize that the embeddings can also identify shared variations and gradients across different counties. By looking at cross-county embeddings through a transferability analysis, we can measure the extent to which the spatial structure of a county shares similar patterns with other counties. The results and findings show that the graph neural network (GNN)-based framework enables us to integrate multiple features in urban areas and their non-linear relationships to understand the underlying complex interactions among population activities, geography, facility services, and socio-demographic contexts that shape spatial structure and gradient of cities.

\section*{\sffamily Methods}
\vspace{-0.5em}
\subsection*{\sffamily Data collection and processing}
\vspace{-0.7em}
This study focuses on 16 metropolitan counties in the United States. We select these counties based on several criteria: (1) the counties would rank among the top 50 US counties in population; (2) the selected counties should be distributed in different states and regions of the United States; (3) facility and mobile phone data should be available; and (4) no extreme event or crisis significantly influencing the activities of populations during the study period had occurred in selected counties. By considering these criteria, we selected 16 counties: Cook County (Chicago, Illinois), Harris County (Houston, Texas), Dallas County (Dallas, Texas), Fulton County (Atlanta, Georgia), King County (Seattle, Washington), Wayne County (Detroit, Michigan), Hennepin County (Minneapolis, Minnesota), Allegheny County (Pittsburgh, Pennsylvania), Miami–Dade County (Miami, Florida), Multnomah County (Portland, Oregon), Sacramento County (Sacramento, California), Salt Lake County (Salt Lake City, Utah), Tarrant County (Fort Worth, Texas), Travis County (Austin, Texas), Washington County (Portland, Oregon), and Santa Clara County (San Jose, California). We used multiple data sets from different sources to generate features for urban areas (see \textbf{Table \ref{tab:data}}). To ensure consistency across all data sets, we first defined a grid map enabling areas of equal size, and then associated all the features with the grid cells. Details about the data and approach are provided in the following paragraphs.


\textbf{Grid map creation.} The diverse features we extracted for this study are present at different scales. For example, socio-demographic features are collected at the census-tract (CT) level, while facility service can be finer due to detailed geographical coordinates provided for the POIs. To explore the spatial structure based on a consistent and fine scale, we divided the area of a county into grid cells of relatively equal size, using equal division of the longitude and latitude spans of a county. The exact size of grid cells might differ among counties when we project the coordinates into distance in counties at different latitudes. For example, Harris County has 4.73 MM population, making it the most populous county in Texas and the third-most populous county in the United States. Considering both the computational cost and fine-scale requirement of analyses, we set a grid cell to be about 4 km x 4 km. The grid cells use the lat-long coordinates of the corner points as the numeric reference to the geographic partition. Features such as facility services are aggregated and counted at the grid level. Socio-demographic features are expressed in percentages. We assume that these features are uniformly distributed in the CT. Hence, we assign the feature values of the census tracts to grid cells within the same CT. 

\textbf{Population estimation.} For this step, we adopted the anonymized mobile phone data from X-Mode, a location intelligence company, which supplies billions of records of anonymized and high-resolution mobile phone location pings for 50MM devices globally. X-Mode works with more than 70 developers of more than 300 applications, who use their proprietary software development kit (SDK) technology \cite{x-mode}. The location data are collected from mobile phone users who opt-in to share their data anonymously through a General Data Protection Regulation and California Consumer Privacy Act compliant framework. In the United States, more than 30 million unique mobile phone devices generate 2 to 3 billion pieces of location data per day. The data was shared under a strict contract with X-Mode through their academic collaborative program in which they provide access to de-identified and privacy-enhanced mobility data for academic research. All researchers processed and analyzed the data under a non-disclosure agreement and were obligated to not share data further and not to attempt to re-identify data.

We filtered the mobile phone data generated during February 2020 in the 16 selected counties. No large-scale extreme event occurred in these counties during this period. We considered population activities in this period as the normal conditions. The home locations of the mobile phone devices are not reported by X-Mode. To estimate the population size at the grid-cell level, referring to prior studies \cite{Moro2021}, we first assigned each data point generated between the hours of 10 p.m. and 7 a.m. to its respective grid cell. The most common grid cell locations of the device during this observation period are considered as the home grid cell. Mobile phones which do not have the location data for more than 2/3 days of the month are discarded.

\textbf{Mobility network construction.} The mobility network is constructed to capture the spatial connections among different urban areas using population flows. In the mobility network $\mathcal{G=(V,E,W)}$, we consider the nodes ($v \in \mathcal{V}$) to be grid cells, and the links ($e_{ij} \in \mathcal{E}$) are the movements from one grid cell to another. The network is weighted and directed, meaning that each link is directed from the origin of the movement to the destination, and the number of people moving on the same link is considered as the weight of the link ($w_{ij} \in \mathcal{W}$). The weights of the links are calculated using the average number of people moving on the links on the day of a week during February. The features related to network characteristics in \textbf{Table \ref{tab:data}} are computed using this mobility network. The in-degree of a grid cell is defined by the number of links directed to the grid cell without considering the weights of the links. The weighted in-degree is the sum of the weights of the links directed to the grid cell. These network features capture the popularity and attractiveness of a grid cell in terms of population activities.

\begin{table}[ht]
\centering
\caption{Selection of features of urban areas for a county}
\label{tab:data}
\resizebox{\textwidth}{!}{%
\begin{tabular}{c|p{0.28\linewidth}p{0.28\linewidth}|c} \hline 
\textbf{Category} & \multicolumn{2}{c}{\textbf{Feature details}} & \textbf{Source} \\ [6pt]
\hline
\multirow{2}{*}{\textbf{Socio-demographics}} & Percentage of minority & Per capita income & \multirow{2}{*}{U.S. Census} \\ [2pt]
 & Percentage of people older than 65 & Percentage of crowded structures &  \\ [2pt] \hline
\multirow{9}{*}{\textbf{Facility services}} & Utilities & Construction & \multirow{9}{*}{Points of Interest (POI)} \\ [2pt] 
 & Manufacturing & Wholesale trade &  \\ [2pt] 
 & Retail trade & Finance and insurance &  \\ [2pt] 
 & Professional, scientific, and technical services & Transportation and warehousing &  \\ [2pt] 
 & Real estate rental and leasing & Information &  \\ [2pt] 
 & Administrative services & Educational services &  \\ [2pt] 
 & Health care and social assistance & Arts, entertainment, and recreation &  \\ [2pt] 
 & Accommodation and food services & Other services (except public administration) &  \\ [2pt] 
 & Public administration &  &  \\ [2pt]  \hline
\textbf{Population} & \multicolumn{2}{c|}{Residential population size} & \multirow{3}{*}{Mobile phone data} \\ [2pt]  \cline{1-3}
\multirow{2}{*}{\textbf{Network characteristics}} & \multicolumn{2}{c|}{In-degree} &  \\ [2pt] 
 & \multicolumn{2}{c|}{Weighted in-degree} & \\ [2pt]  \hline
\end{tabular}%
}
\end{table}

\textbf{Other data.} We obtained the socio-demographic information from the census data publicly available on the website of the U.S. Census Bureau. We used mainly the American Community Survey (ACS) 2014–2018 (5-year) data \cite{UnitedStatesCensusBureau2019} to extract the socio-demographic features in \textbf{Table \ref{tab:data}}. The data are provided at the census tract level. The facility data are obtained from SafeGraph, Inc., a location intelligence data company that has collected the geographical and business information about physical places, also called “points of interest” (POI), in the United States \cite{SafeGraph2020}. POIs are dedicated geographic entities, such as restaurants, retail stores, and grocery stores. SafeGraph POI data include about 6.5 million active POIs with the information of geographical coordinates, industry category code, brand, address and administrative region of the POI. The industry category code of the POI is developed by the North American Industry Classification System (NAICS), which is the standard used by Federal statistical agencies in classifying business establishments \cite{naics}. To have a comprehensive consideration of the POIs, we use the first two digits of the NAICS code, which is the highest-level categorization. The criteria for choosing the types of facility services (i.e., POIs) in this study include: (1) the facilities should provide services for the majority of the population; (2) different types of POIs should provide distinguishable services or serving distinguishable population groups; and (3) sufficient data points should be available for specific types of POIs in a county. The SafeGraph Data Consortium shared the data through their academic program, which provides data at no charge to the academic community. The program allows researchers to publish papers and uncover insights that will lead to better policymaking, innovations, and business growth.

\subsection*{\sffamily Neural Embedding Model}
\begin{figure}
  \centering
  \includegraphics[width=17cm]{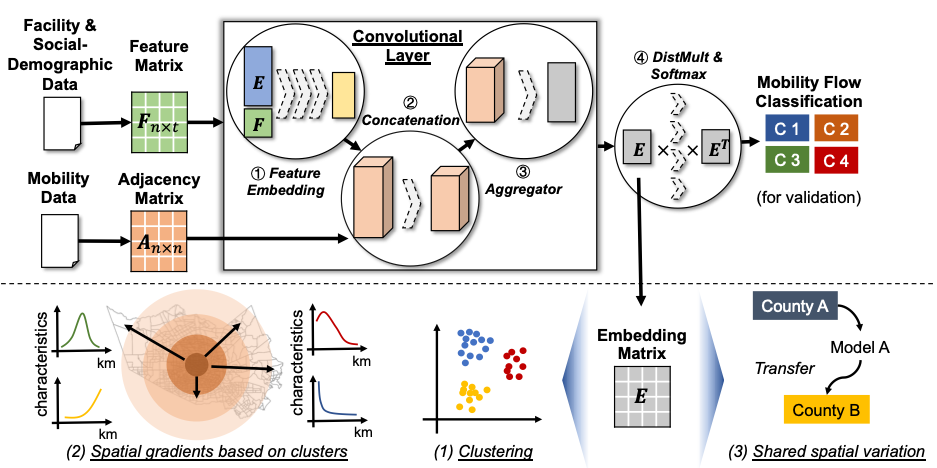}
  \caption{\textbf{Illustration of the analytical framework.} The framework is composed of two components: the architecture of the graph neural network (GNN) and the analyses on generated neural embeddings. GNN takes the adjacency matrix generated from anonymized mobile phone data and the feature matrix. Features for each urban area (grid cell) are income, ethnicity, age and building information from the social-demographic data for census tracts, and the number of specific facilities provided by SafeGraph. The graph convolutional layer can translate the feature and adjacency matrices to encode not only features of the grid cell itself, but also the features of its neighbors through the consideration of the topological structure of each node’s neighborhood as well as the distribution of node features in the neighborhood. The DistMult layer is adopted to integrate the embeddings of two grid cells and make a prediction for the movement flows between these two cells using the SoftMax function. The vector representations of grid cells are in the embedding matrix. We clustered the urban areas using the representations to capture emergent spatial structure and analyze the spatial gradients in a county and the shared spatial variations across different counties.}
  \label{fig:fig1}
\end{figure}

\vspace{-0.7em}
We compile all features for grid cells in a feature matrix, $F=(F_1^T,F_2^T,…,F_{|\mathcal{V}|}^T)$, where the $i^{th}$ element $F_i^T\in\mathbb{R}^d$ is a vector representation including all values of d features for grid cell $v_i$. The adjacency matrix $\mathcal{A} \in \mathbb{R}^{|\mathcal{V}| \times |\mathcal{V}|}$ of the urban areas is created based on the mobility network $\mathcal{G}$. We consider two urban areas to have connections when there is at least one person moving between these areas during the study period. With these two matrixes, we adopted the graph convolutional neural network model to encode the information and relationships among urban areas. The model is used to seek effective representations of urban areas that are useful for revealing the structure of cities. The GNN model includes two comprehensive computational components in an iterative process (see \textbf{Fig. \ref{fig:fig1}}): graph convolutional layer, which is used to encode both feature matrix and adjacency matrix to reveal representative embeddings of urban areas through stacked layers of dense networks; and DistMult layer, which is used to classify the links using the embeddings of the starting and ending nodes. 

\textbf{Graph convolutional layer.} The output of this layer is the embeddings of urban areas, which are generated through three components of the layer. First, the feature embedding component stacks three layers: linear layer (a fully connected network), normalization, and drop out. This component initializes the embedding matrix $E \in \mathbb{R}^{|\mathcal{V}| \times d}$ first and combines it with the feature matrix. Through dimension reduction, the two matrixes will result in one hidden matrix with $d$ dimensions. The second component is concatenation, which operates the hidden matrix to pass the information from the neighbors of a node based on the adjacency matrix. Then the aggregator is used to aggregate the information from the neighbors of a node. Here, we use a non-parametric function, simple averaging embeddings of all nodes in the neighborhood to construct the neighborhood embedding. The average value is passed through a fully connected layer. In this graph convolutional layer, the information propagation \cite{schlichtkrull2018modeling} from the previous layer to the current layer is:

\begin{equation}
    h_i^{(l+1)} = \sigma (\sum_{e \in \mathcal{E}} \sum_{j \in \mathcal{N}_i^e} \frac{1}{c_{i,e}} W_e^{(l)}h_j^{(l)} + W_0^{(l)}h_i^{(l)})
\end{equation}

where $h_i^{(l)} \in \mathbb{R}^{d^{(l)}}$ is the hidden vector representation of node $v_i$, in the $l$-th layer of the neural network, with $d^{(l)}$ being the dimensionality of this layer’s representations; $\sigma$ is an activation function ($\tanh()$ in this study); $\mathcal{N}_i^e$ denotes the set of neighbor indices of node $i$ with the link $e \in \mathcal{E}$; $c_{i,e}$ is the normalization constant $(c_{i,e}=|\mathcal{N}_i^e |)$; $W$ is a learnable weight matrix. This function passes the hidden information from the neighbors of a node based on the directed link from the origin to the destination. 

\textbf{DistMult layer.} A DistMult decoder that takes pairs of embeddings of urban areas and produces a score for every link with weights in the mobility network. The loss is evaluated per link. The DistMult factorization in this study is adopted from the technique by Yang et al. \cite{Yang2015EmbeddingEA}, which is a scoring function. In the DistMult layer, any pair of area embeddings is scored as:

\begin{equation}
    f(i,j) = h_i^T M h_j
\end{equation}

where $h_i$ is the embedding of the urban area $i$ where a link originates, $h_j$ is the embedding of the urban area $j$ where a link directs to, and $M$ is a learnable diagonal matrix. We calculate the score of the links for each class. That is, we applied four linear layers to manipulate the two embeddings of the origin and destination. 

To validate the embeddings of the GNN model, we conducted a supervised learning approach to capture human mobility in urban areas. In this task, we divided the links into classes in terms of volumes of the movements. Since the data usually are imbalanced, the classes tend to be imbalanced. That is, the number of links in some classes is extremely high or extremely low. To take into account the issues of imbalanced data, we adopted a weighted cross-entropy loss function: 

\begin{equation}
    \mathcal{L} = -\frac{1}{\sum_{k=1}^K w_k} \sum_{(i,j) \in \mathcal{T}} \sum_{k=1}^K w_k \cdot t_{ij,k} \cdot \log \Theta(f_k(i,j))
\end{equation}

where $w_k$ is the weight of a class, which is calculated based on the rate between the number of samples in the first class and the number of samples in other class, $w_k=N_1/N_k$. Here, the first class is the largest class in the data set. $t_{ij,k}$ is a binary variable, corresponding the respective ground truth class $k$ for link $e_{ij}$. If the link is in class $k$, then $t_{ij,k}$ equals 1; otherwise, $t_{ij,k}$ equals 0. $\Theta$ is an activation function. In this step, we used SoftMax on the output of the last layer from DistMult. We minimized the resulting cross-entropy loss on all present edges. $f_k (i,j)$ is the score of the link $e_{ij}$ for the class $k$. $K$ is the total number of classes in this classification task. $\mathcal{T}$ is the total set of pairs of urban areas. 

\textbf{Hyperparameter tuning.} To improve its efficiency in training and accuracy of prediction, we tuned a few hyperparameters of the model: learning rate (0.025, 0.05), the number of dimensions (50, 75, 100, 150). The learning rate is tuned to obtain an appropriate converging speed for the algorithm. Since this downstream task is implemented to obtain valid embeddings of urban areas, the highest accuracy of prediction outcomes is not particularly required. Hence, we set the learning rate to be 0.05 in training models for all counties, reducing the time for model training but retaining the quite good quality of the model. In addition, the number of dimensions of the vector representations matters for the quality of the encoding. Typically, the more dimensions the embedding has, the greater the quality of encoding, but, there will also be some limits beyond which diminishing returns will be obtained. We set the minimum dimension to 50 and tested the prediction results on the dimensions of 75, 100, and 150. The results show that embeddings of 100 dimensions could achieve the best prediction accuracy of the downstream task among the results from other dimension options.

\subsection*{\sffamily Clustering and cross-county embeddings}
\vspace{-0.7em}
\textbf{Agglomerative clustering.} The neural network embeddings generated from the GNN model encode all the attributes and topological structures of the urban areas. The complex relationships among urban components and population activities and attributes in urban areas can be captured through clustering the representations which are close to each other. In this task, we employed agglomerative clustering \cite{Murtagh2014}, a hierarchical clustering using a bottom-up approach. This approach first considers that each embedding starts in its own cluster, and then the pair of clusters would be recursively merged when an increase in a given linkage distance is minimum. We used the Euclidean distance to find which clusters to merge, using the algorithm proposed by Ward \cite{Ward1963}. Agglomerative clustering is a powerful technique that allows us to build tree structures from data similarities; it is particularly useful in network data. Our embeddings encode the spatial connectivity and topology of the spatial structure. Hence, the agglomerative clustering approach is a proper approach to cluster the urban areas based on their vector representations.

\textbf{Cross-county embeddings.} To test the similarities of spatial structure among different counties, we conducted cross-county experiments. First, we trained the GNN model by feeding the model with the feature and adjacency matrixes from county A. After the model reached an acceptable prediction accuracy, we saved the model with fixed parameters. We then input the adjacency and feature matrixes from county B to generate the embeddings for the urban areas in county B. The transferrable model would output similar embeddings for urban areas in county B, when these areas are assigned the same cluster using representations from the model trained on county B. That is, two counties share a highly similar spatial structure if the majority of the pairs of urban areas are in the same clusters, even the models are trained on different counties. Hence, we measure the similarity of spatial structure in different counties through the percentage of area pairs in the same cluster.

\section*{\sffamily Results}
\vspace{-0.5em}
\subsection*{\sffamily Generating and validating the embedding space}
\begin{figure}
  \centering
  \includegraphics[width=17cm]{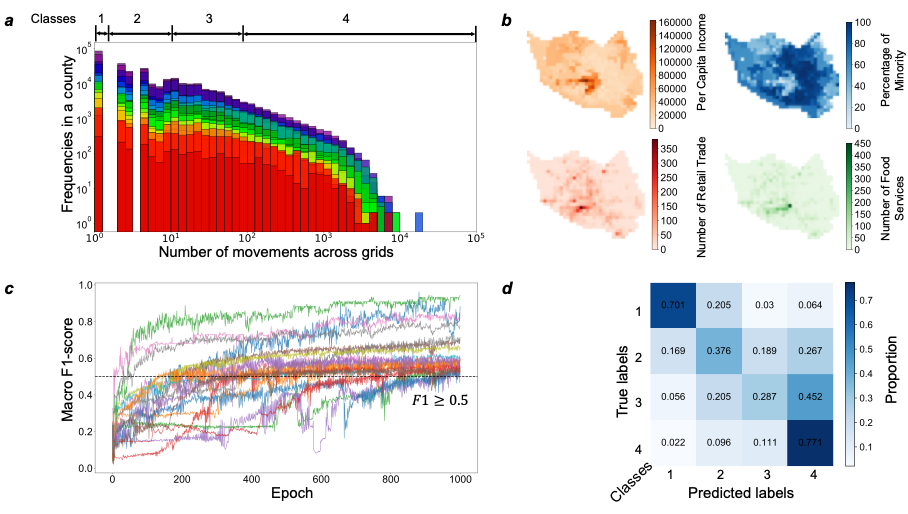}
  \caption{\textbf{Input data and the performance of model training.} {\sffamily \textbf{a.}} A histogram showing the distribution of movement flow across grid cells in a county. The distributions for all the counties selected in this study are plotted in the figure (represented by different colors). Based on the distributions of flows in the log-log scale, we divided the movement links into four classes: $\leq 10^0$; $10^0<$ and $\leq 10^1$; $10^1<$ and $\leq 10^2$; and $>10^2$. {\sffamily \textbf{b.}} geographical maps of four features: per capita income; percentage of minority population; number of retail trade establishments; and the number of food services, such as restaurants, in Harris County as an illustrative example. {\sffamily \textbf{c.}} Macro F1-score as the metric of the performance of the model in predicting the classes of movement flows, versus the number of training epochs. We defined the threshold of 0.5 for the F1 score in to obtain compact embeddings of urban areas. {\sffamily \textbf{d.}} Confusion matrix indicating the prediction performance of the GNN model for movements in Harris County. Values in each row sum to 1.}
  \label{fig:fig2}
\end{figure}

\vspace{-0.7em}
Before implementing the model and generating the embeddings, we first explored the observable patterns of urban attributes in a county. In this section, we take an example of Harris County and show the spatial distributions of some socio-demographic and facility attributes (\textbf{Fig. \ref{fig:fig2}b}). As we see in the \textbf{Fig. \ref{fig:fig2}}, people with high income usually concentrate in a very small area tending to be in the center of the county, surrounded by a low-income population in the majority of the urban areas. The suburban areas, such the areas in North of Harris County, however, also have high-income populations. In addition, minority populations are shown to live outside the center of the county, but they do not concentrate in the marginal area of the county. It is obvious that the income and percentage of the minority are not distributed on a linear gradient from the center of the county to the surrounding territory. The other two maps in \textbf{Fig. \ref{fig:fig2}b} show the spatial distributions of facility services (retail trade and food services), which are different from the distributions of the socio-demographic attributes. We can find that the majority of the facilities locate at the center of the county, and the number of facilities decreases with distance from the center to the margin. The gradients of these facilities, however, are not uniform. For example, the gradient to the southwest of Harris County is small, while the gradient to the east is quite sharp. These observations show us a multifaceted urban structure with highly unequal distributions of urban attributes. It is also difficult to conclude a clear mathematical expression to describe the relationships between urban areas and how spatial gradients emerge in different directions from the center to the margin. Such complex relationships of areas in multifaceted urban systems are responsible for challenges in the understanding of emergent spatial structures of cities. This challenge raises the necessity and importance of advanced models to capture the non-linear relationships among urban areas and heterogeneous features related to urban components and population activities. 

To this end, we calibrate the GNN model to integrate features from distinct aspects of urban areas and generated effective representations to characterize the emergent spatial structures of a county. To validate the embeddings generated from the GNN model, in this step, we designed a downstream task to enable supervised learning so that we can update the embeddings by minimizing the loss of information. The downstream task is a classification of links’ weights (i.e., the number of people moving on the link), since multiple prior studies have considered human mobility to be highly influenced by urban spatial structures \cite{Bassolas2019,DiClemente2018}. In this task, the input of the model is the feature and adjacency matrixes of the urban areas. The DistMult layer projects the embeddings of a pair of urban areas, obtained from graph convolutional layer, into a score representing the strength of the relationship between two urban areas. A SoftMax function is then used to classify the score into different classes. \textbf{Fig. \ref{fig:fig2}a} shows the distributions of movement volumes on links in the mobility networks of each county and how we create the classes. The distribution is plotted in a log-log scale since it follows a long-tailed distribution; that is, the majority of the links have very few movements, while a few links have a very large volume of traffic. The slope for the distribution is quite steep, indicating a highly imbalanced distribution. Such patterns can be observed in all distributions for different counties. Hence, to mitigate the imbalance, we divide the range of movement volumes based on the exponent of 10. By doing so, we have four classes for the movements on links, which are consistent for all selected counties in this study. This setting will also be beneficial for the following cross-county embedding and analyses. 

With the labels and inputs, we trained the models with all node features and complete adjacency matrix, but randomly sample 80\% of the links for the supervised learning of the downstream task. \textbf{Fig. \ref{fig:fig2}c} shows the training process and the extent to which the performance of the models is improved for all counties. We used the macro F1 score to measure the performance of the model, since the macro F1 score allows us to equally consider the importance of classes and the classification accuracy. The figure shows that the models in different counties converge at different speeds. The F1 scores on 1,000 epochs for different counties are also significantly different. For example, F1 scores for models of a few counties can reach 0.8 in less than 100 epochs, while models of some counties have to take more than 800 epochs to reach F1 of 0.5. The variations of converging speed may be due to the size of the county and the relationship between human mobility and the spatial structure of a county. Since this task is only for validating the embeddings, we do not require an extremely high F1 score. Hence, to save time on training models, we set both thresholds for the number of epochs in training and the F1 score. informed by existing studies \cite{Peng2021,Xue2021QuantifyingSH}, we would accept the validity of the embeddings when the F1 score of the model is greater than 0.5. In this study, the performance of the models for all selected counties satisfies this criterion. Hence, the embeddings are valid in all these counties. \textbf{Fig. \ref{fig:fig2}d} shows an example of the classification results in Harris County using a confusion matrix. The model achieves very good performance especially in classes 1 and 4 with an F1 of more than 0.7. Although the results in classes 2 and 3 are not as good as the other classes, the model can still capture the level of movement volumes for a large portion of links. The model reaches a very low probability to assign high-traffic links to class 1 where the links have just one movement. Based on these observations, we could accept that the model can capture human mobility in urban areas and the embeddings generated from the model are valid for characterization of emergent spatial structures and gradients in cities.

\subsection*{\sffamily Spatial structure revealed by embeddings}
\begin{figure}
  \centering
  \includegraphics[width=17cm]{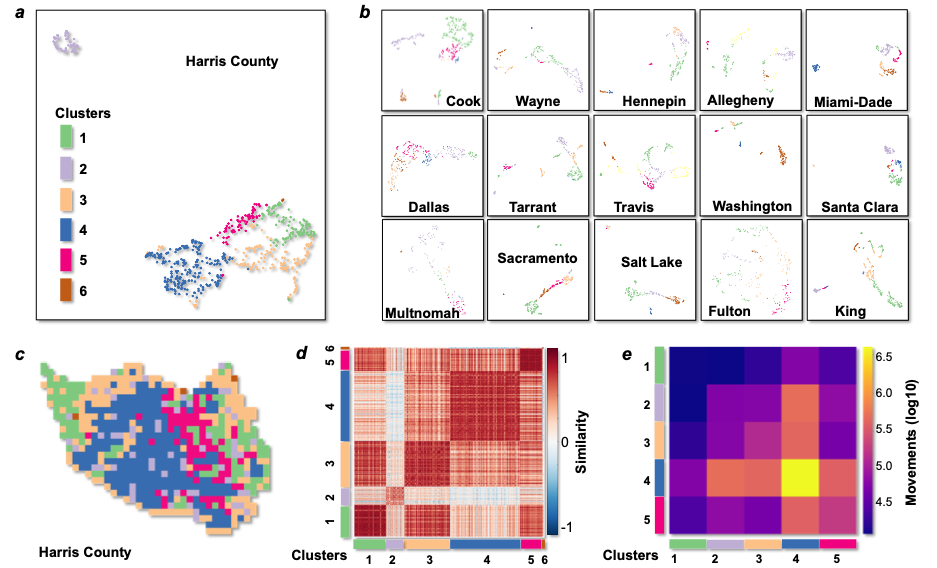}
  \caption{\textbf{Embeddings of urban areas reveal multifaceted spatial structures of counties.} {\sffamily \textbf{a.}} The two-dimensional (2D) projection of grid cells in Harris County obtained from dimension reduction of GNN embedding matrix using UMAP approach \cite{McInnes2018UMAPUM}. Each dot represents a grid cell (urban area), and its color denotes its cluster designated by the clustering method. {\sffamily \textbf{b.}} The 2D projection of grid cells in the rest 14 studied counties. Different colors represent different clusters. {\sffamily \textbf{c.}} Geographical map of Harris County with color-coded grid cells. The colors for clusters are consistent with the colors defined in (a). This map shows the spatial distribution of the clustered urban areas based on GNN embeddings. {\sffamily \textbf{d.}} Similarities between grid cells calculated using cosine similarity and the vector representations. The scores of similarities span from -1 to 1. {\sffamily \textbf{e.}} Flow of movement between grid cells in different clusters. The cluster 6 contains very few grid cells, and the movement flows from and to this cluster is very low. Hence, it is excluded from this matrix. The rows represent the origin from which the movements come, and the columns represent the destinations where the movements travel to.}
  \label{fig:fig3}
\end{figure}

\vspace{-0.7em}
Once the GNN model is calibrated for each county, we can obtain the embeddings for urban areas. The initial output of the embeddings had 100 dimensions (see Methods section) which encoded the features of the urban area itself as well as complex structure in the high-dimensional space. If our embeddings are indeed capable of capturing the complex relationships among urban areas, it is reasonable to hypothesize that certain areas would have strong connections and very short distances in the high-dimensional space. Areas that are distinct from other areas should also be distant from each other in the high-dimensional space. To test this hypothesis, we clustered the urban areas using agglomerative clustering. The example of Harris County, \textbf{Fig. \ref{fig:fig3}d} shows the pair-wise cosine similarities of the embeddings to validate the results of clustering. The results are displayed in the order of clusters. We compared the similarity matrixes of inner-cluster embeddings and between-cluster embeddings. We can observe that the components of the similarity matrix on the diagonal tend to be darkest red, compared to the other components. Although some embeddings like the embeddings in cluster 3 also have a relatively strong agreement with the embeddings in other clusters, the similarities are still weaker than the similarities within the cluster. Hence, these results can still be a strong evidence that inner-cluster similarities are higher than between-cluster similarities, validating the outcomes of the agglomerative clustering. 

To provide an intuitive visualization of the hidden structures in the embeddings, we employed the UMAP (Uniform Manifold Approximation and Projection for Dimension Reduction) approach \cite{McInnes2018UMAPUM} to reduce the dimensionality of embeddings. As demonstrated in prior studies, UMAP can better display results with closer inner-cluster distances and larger between-cluster distances \cite{Wang2021}. \textbf{Fig. \ref{fig:fig3}a} shows a two-dimensional (2D) projection and visualization of the embeddings of urban areas in Harris County, providing an overview of the emergent structure in the county. In this figure, we have six clusters of urban areas in Harris County. Since there are very few nodes in cluster 6, its evidence is insufficient to pinpoint the nodes in cluster 6 from the projection. For other clusters, however, the result clearly shows the closeness and nuanced structure of urban areas with conspicuous regions in our projection. At the individual node level, we can simply distinguish the nodes from different clusters since the nodes in the same cluster locate close to each other. There is little overlap between nodes from different clusters. At the cluster level, the distances between different clusters vary. In the case of Harris County, we observe that cluster 2 is far from other clusters, implying significant differences with regard to the features and human mobility patterns. Meanwhile, the number of nodes in a cluster and the distribution of the node in this two-dimensional space vary across clusters. For instance, nodes in cluster 2 are more concentrated, while nodes in cluster 3 are sparsely distribute in the space. Although the specific locations—x and y axis in the projection, for instance—are meaningless and unexplainable, the degree of accumulation and dispersion can reflect the nuanced relationships among urban areas. These patterns can also be observed in other counties but with variations across the counties (\textbf{Fig. \ref{fig:fig3}b}). In Miami–Dade County, similar to Harris County, one cluster of the urban areas are distant from other clusters in the embedding space, while urban areas have loose relationships with each other in Fulton County. Such differences motivated us to explore the underlying mechanisms responsible for the heterogeneity of urban areas in the high-dimensional embedding space. 

We further look into the spatial patterns of the clusters in the physical space by color-coding grid cells based on clusters. \textbf{Fig. \ref{fig:fig3}c} shows the example of Harris County. Areas in cluster 4 are primarily in the center of the county, surrounding areas in cluster 2. In general, areas in cluster 3 disperse at the margin of the county, while areas in cluster 5 concentrate in a small region. The changes of clusters from the center to the margin of the county vary in different directions. For example, from the center to the north and south, there are only two: clusters 4 and 3. From the center of Harris County and progressing east and southeast, the clusters are 4, then 5, then 1. From the center moving northwest, clusters change from 4, to 3, and finally to 1. This result indicates the anisotropic spatial gradient in a county. Prior studies \cite{Liu2018vertical} tend to use the distance from the center to the margin and consider isotropic spatial gradient, which is not sufficient to understand the complex and emergent spatial structure of counties. This finding shows that dense urban area embedding is a promising data-driven approach to effectively analyze spatial gradients in complex urban space using vector similarity. Meanwhile, the findings raise important questions regarding the mechanisms of why the emergent spatial structure shape in this way and what specific characteristics urban areas in the same cluster share. 

\subsection*{\sffamily Spatial gradients in multifaceted emergent structure}
\vspace{-0.7em}
The embeddings of urban areas encode a number of heterogeneous features and the complex interactions between urban areas in the high-dimensional space. The spatial distribution patterns in both high-dimensional and physical spaces motivated us to explore the underlying mechanisms regarding the organization of the structures. With clustered urban areas using embeddings, we could reveal the multifaceted emergent structure of a county. In this section, we summarize the characteristics of the areas in each cluster of Harris County to explain the spatial patterns we previously observed. We first examine population movements in the county, which reflect the spatial connections among different urban areas. \textbf{Fig. \ref{fig:fig3}e} shows the within- and across-cluster movements in Harris County. We can find that areas in cluster 4, which are at the center of the county, account for a great number of in-flow and out-flow movements. Areas in the margin of the county, especially areas in cluster 1, are marginalized in terms of human mobility, indicating limited interactions between these areas and the center of the county. Overall, the mobility network based on clusters presents a hierarchy. The majority of the movements occur between the areas in cluster 4. Then, the movements flow to areas in clusters 2, 3 and 5. The weakest level of movements is the movements between areas in cluster 1 and other areas. These findings show the interactions between human mobility and the spatial structure of the county, which are captured by the high-dimensional embeddings.

\begin{figure}
  \centering
  \includegraphics[width=17cm]{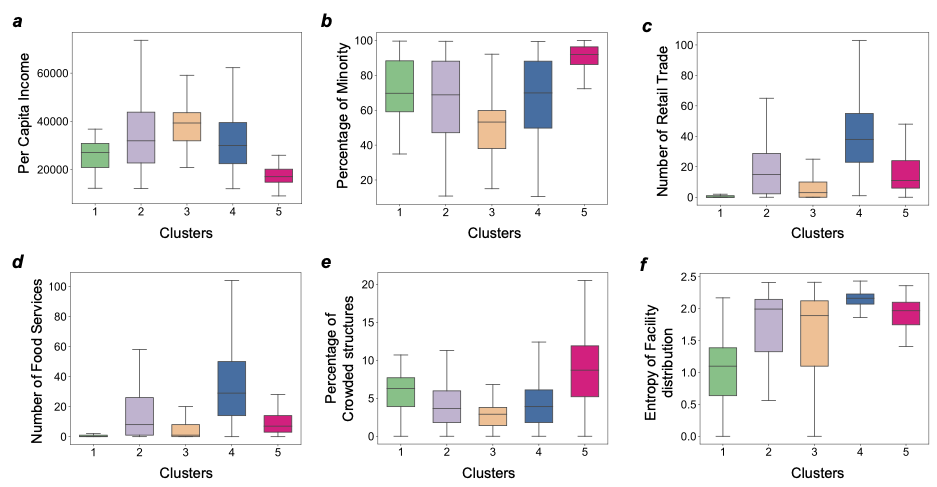}
  \caption{\textbf{Characterization of clusters and the spatial gradients for Harris County.} These analyses only include five clusters since cluster 6 is too small with very few grid cells. The features as examples to study the spatial gradients include: {\sffamily \textbf{a.}} per capita income; {\sffamily \textbf{b.}} percentage of minority population; {\sffamily \textbf{c.}} number of retail trade points of interest; {\sffamily \textbf{d.}} number of food services; {\sffamily \textbf{e.}} percentage of crowded structures; {\sffamily \textbf{e.}} entropy of facility distribution. Results for other counties can be found in the Supplementary Information.}
  \label{fig:fig4}
\end{figure}

\textbf{Fig. \ref{fig:fig4}} is a summary of typical features of urban areas in five major clusters. We can find that people living in areas in cluster 3 tend to have the highest income, while people living in areas of cluster 5 may be in the lowest-income population group (\textbf{Fig. \ref{fig:fig4}a}). Areas in cluster 3 usually account for a very low proportion of minorities, while the majority of the people living in area 5 are minorities (\textbf{Fig. \ref{fig:fig4}b}). This result indicates the presence of income and race segregation between cluster 3 and other clusters. Other areas like areas in clusters 1, 2, and 4 are in the middle in terms of per capita income and percentage of minorities. Regarding facilities, we can observe that areas in cluster 4 have a greater number of facilities, such as retail trade and food services. Areas in cluster 1, however, have very few facilities. These areas may only be residential areas (\textbf{Fig. \ref{fig:fig4}c} and \textbf{\ref{fig:fig4}d}). To further explore what types of buildings that people inhabit in these areas, we provide information for the percentage of crowded structures. We find that areas in cluster 3 have the lowest percentage of crowded structures, while areas in cluster 5 tend to have the highest density of crowded structures (\textbf{Fig. \ref{fig:fig4}e}). In addition, areas in cluster 4 have the highest entropy for facility services (\textbf{Fig. \ref{fig:fig4}f}), indicating the diversity of facilities that can satisfy the demand of the population on different life activities. Since areas in cluster 1 do not have too many facilities, the diversity of facilities in areas of cluster 1 is also the lowest among the five major clusters. This result shows disparate patterns of facility distribution in different clusters. 

Considering together the findings from spatial patterns and characteristics of the clusters, we can infer underlying mechanisms of how the multifaceted emergent structures are spatially formed. First, the center of Harris County has a large and diverse number of facilities. The income of persons living in these areas, however, is not among the highest. People with the highest income may live in the marginal areas of the county with a very low proportion of minority population, crowded structures, and fewer facilities, compared to people living in the center of the county. The neural network embeddings also capture a small region, areas in cluster 5, where low-income people are Low-income minorities live in crowded structures, although they have moderate number of and a diverse of facilities. We find a clear transition of the features from the center to the margin of the county, but it is anisotropic in different directions. The diversity of facilities may decrease along with the distance from the center, but the socio-demographic attributes present a different trend. For instance, from the center to the east of Harris, the percentage of minority goes up and reaches a peak in areas of cluster 5, and then decreases to its lowest point in areas of cluster 3 at the margin of the county. Per capita income increases moving from the center to the northwest in areas of cluster 3 and decreases in areas of cluster 1. These non-linear patterns may not be captured by prior studies which focus only on the simple and linear structures of cities.

\subsection*{\sffamily Cross-county embeddings for shared spatial structures}
\vspace{-0.7em}
Another capability of the neural network embedding model is to examine the similarities of spatial structures across cities. It is evident that, in physical space, the attributes and structures of different cities are distinct. However, this might not be the case in the high-dimensional space, which captures complex and emergent structures of the cities. Identifying similarities of cities’ spatial structures is important for making policies adaptive to broader regions and expanding lessons learned from one city to other cities. The capability of the embedding model prompted us to explore the general patterns and understand the degree of similarities among the cities in the high-dimensional space. 

\begin{figure}
  \centering
  \includegraphics[width=17cm]{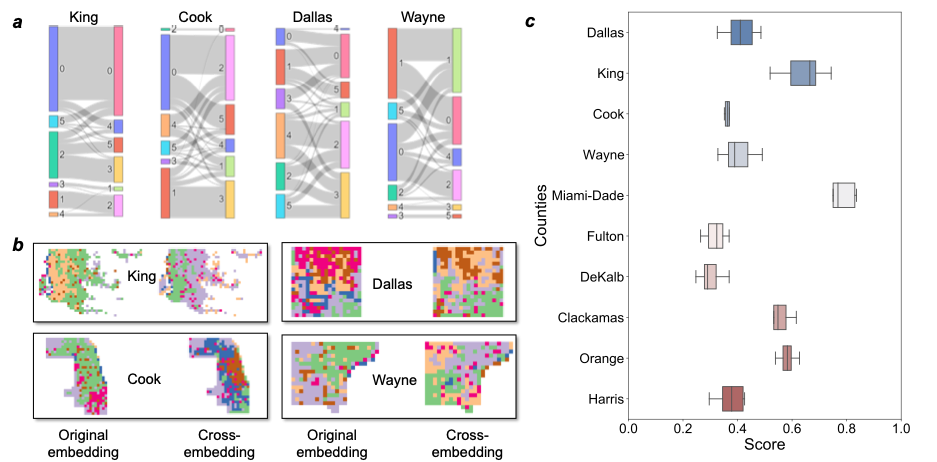}
  \caption{\textbf{Analyses of cross-county embeddings to measure shared spatial structures among different counties.} {\sffamily \textbf{a.}} The left column is the clusters based on the embeddings generated from the GNN model trained on the same county. The right column is the clusters based on the embeddings generated from the model trained on Harris County. We show the results for four example counties using the model trained on Harris County. {\sffamily \textbf{b.}} Geographical visualizations of the classified grid cells in each county using different models. Colors do not correspond to counties in the same box. {\sffamily \textbf{c.}} The percentage (“score” in the chart) of pairs of grid cells in the same cluster using original and cross-county models. We conducted pair-wise experiments for these ten counties. That is, the model trained on one county generates embeddings for the rest of the counties. Each box in the chart shows the average and variation of ten scores.}
  \label{fig:fig5}
\end{figure}

We first trained the embedding model on a county, such as Harris County in the example case (\textbf{Fig. \ref{fig:fig5}a}). Once the classification performance (F1 score) of the model satisfies our criterion, we apply the model to generate embeddings for another county by feeding the feature and adjacency matrixes of that county. We call these embeddings cross-county embeddings. Then, we cluster the embeddings using the same clustering methods. The similarities of spatial structures are defined as the extent to which the complex relationships between two urban areas captured in original embeddings (embeddings generated from the model trained on itself) can also be captured in cross-county embeddings. In sum, the pair of urban areas can be assigned to the same cluster using cross-county embeddings as that using original embeddings. Based on this definition, by implementing the cross-county embedding approach, we can find that the majority of the pairs of areas are assigned in the same clusters in King County, especially for areas in clusters 0, 1, and 2. This result indicates that the spatial structure of Harris County is particularly similar to the structure of King County, since the patterns learned from Harris County can also be applied to cluster the areas in King County. This finding can also be observed in Cook County and Wayne County. However, the spatial structure of Dallas County does not share many similarities with the structure of Harris County since the clusters of the areas change considerably from the results of the original embeddings to the results of cross-county embeddings. These results show that the cross-county embeddings could provide a comparison regarding the extent of similarity and differences among emergent spatial structures of various counties.

We also examined the geographical distributions of the clusters in these cases (\textbf{Fig. \ref{fig:fig5}b}). Although a few areas are assigned into different clusters, the spatial gradients in the structures remain relatively consistent, especially for King, Cook, and Wayne counties. For instance, the center of Cook County is in the mid-east, and the gradients expand from the center to the north, the west and the south. In King County, two clusters in the middle and the east account for a large proportion of the county, and areas in other clusters are dispersed in the west. The case in Dallas County is a bit different, where the cross-county embeddings based on Harris County are not sufficient to capture the geographical distributions of the clusters in Dallas County. That means, it would be challenging to expand computational models or public policies from Harris to Dallas since their spatial structures have quite large differences. These results also motivate us to gain a sense of the extent to which different counties share similar spatial structures. To this end, we conducted experiments of cross-county embeddings for ten counties, including additional counties from different states. We trained ten models on these ten counties, and also generated ten sets of cross-county embeddings for each county (\textbf{Fig. \ref{fig:fig5}c}). Through measuring the pair of areas in the same cluster, it is notable that cross-county embeddings can potentially capture 50\% to 60\% of complex relationships among urban areas in a county. In other words, cities have 50\% to 60\% similarities in their emergent spatial structures. Models trained on counties like Miami-Dade County share a great number of similarities with the structures of other counties, while models trained on Dekalb County can only capture about 30\% of relationships in other counties. These results would also be influenced by the selection of counties, and might change when we include more counties. But, to some degree, the results demonstrate the capability of the proposed method to explore cross-county similarities in examining emergent structures in the high-dimensional space.

\section*{\sffamily Discussion and Concluding Remarks}
\vspace{-1em}
It is still a fundamental challenge to explore spatial heterogeneity in complex urban systems, where socio-demographics, geography, facility services, and population activities intertwine. In this study, we calibrate a GNN-based model to uncover the spatial gradients in multifaceted emergent structures of 16 selected metropolitan counties in the United States. The nature of the spatial structures of cities is emergent and multifaceted with complex and non-linear relationships among features of urban components and population activities and attributes; hence, the previous knowledge and methods may lack the sophistication to accurately characterize urban systems. The key innovations of the proposed method and experiments are integrating the individual and topological features of urban areas using graph neural networks. The benefits of GNN are its intrinsic learnable properties of propagating and aggregating features to capture the nuanced relationships across the whole of urban spatial structures. Hence, the embeddings generated by the GNN model can be treated as the high-order representations of each urban area. Unlike the previous methods and applications in characterizing the spatial structure of cities, which capture only the spatial distribution of single dimension of cities, this study shows the powerful capabilities of neural embeddings to effectively uncover the multifaceted emergent structures of cities. We observed a variety of important findings derived from our experiments and analyses which contribute to the knowledge and practices of the research field.

First, the GNN-based method efficiently captures the complex relationships among urban areas without prior knowledge and assumption of those relationships. The proposed model is a hypothesis-free deep learning framework. Different from previous mathematical or statistical approaches, which require specifying the form of the relations between variables, the neural embedding approach is quite flexible with input variables on data-driven graphs. By constructing the general conceptual dimensions in high-order space, the neural embeddings can expand the capabilities of prior methods. In particular, since the urban areas are numerically represented, we can make analogies between urban areas using the embeddings. This process would allow us to capture how the composition and interactions of urban areas can reflect the status of other urban areas. One typical example is to infer possible outcomes when strengthening or weakening the interactions between two distinct urban areas. These outcomes can be obtained by changing the input values of the features in the GNN model. Existing urban structures with limited capability of encoding the non-linear relationships of the features would fail adapt to the changes. The analogies between urban areas using vector representations could also contribute to derive new knowledge about urban structures as well as new policies for urban development. 

Second, the findings in this study also show the extent to which emergent spatial structures of cities capture various phenomena, such as segregation, disparate facility distribution, and mobility. Concern with societal challenges, including social inequality in cities, has been gaining momentum in both scholarly research and real-life practice. Precisely quantifying and comprehensively understanding social inequalities are particularly important for decision-making and planning strategies. Prior studies may measure inequality without considering the effects of different urban and population features and the interactions of other urban areas. Inequality is a regional challenge under the collective effects of various factors. Evaluating and quantifying inequalities through the integration of features and topologies would be essential. The approach and findings in this study provide a new perspective to dig into the inequality problem by characterizing clusters of urban areas. The results demonstrate the capability of neural embeddings in explaining the gradients present in city expansion and how the urban areas are composited. This outcome could also inform future research to develop a neural embedding-based approach for evaluating social inequalities. Communities could also put the approach and findings into practice to advise urban planning strategies and operation policies. 

Finally, cross-county analyses show that the neural embedding model, to some extent, is transferrable between counties, implying shared spatial structural patterns identified among the counties. The cross-county embeddings provide a metric to quantify the similarities of multifaceted emergent structures among different cities. Previous studies focus only on mobility networks or built environmental features, and still lack quantitative metrics to compare two spatial structures. Current urban polices like the containment of COVID-19 are usually executed at the state level. Human and built environmental patterns in different cities, however, are different, even for the cities in the same state. Specifying the extent to which certain policies can be generalized to other cities and regions is particularly important. Without a proper metric to describe the similarities of the cities, the generalization of policies may be hampered or misleading. In the high-dimensional space, the similarities of spatial structure can be quantitatively measured. The results and findings of these cross-county analyses can contribute to expanding urban policies and strategies from one county to broader regions, which would also be beneficial to improve the effectiveness of policies. For example, containment or urban renewal policies in similar urban areas captured by embeddings in high-dimensional space should be effective, compared to similar areas derived from existing studies without considering complex interactions of urban components. 

Some limitations can still be found in this study, which needs future research to overcome. One important limitation of the neural embedding method is the interpretability of the model and the identified relationships among the urban areas. Since the model does not formulate the relations, it would be challenging to explain how the features and the topologies are integrated into the embeddings. Future research could look into the interpretability of the model with attention mechanisms and graph transformer models which emerge from recent advances in deep learning. Another limitation of this study is the requirements for data sets. As we see in the experiments, we employed a great and diverse number of datasets related to various aspects of the cities. This is because the performance of the neural embedding model is prone to achieve better results with large and comprehensive data sets, compared to relatively small and sparse data sets. Cities with few people using smart devices and less available digitalized facilities information may suffer difficulties in precisely modeling their multifaceted emergent structures and subsequently formulating the spatial gradients. Future research could expand the capabilities and applications of transfer learning in the context of urban studies to enable modeling spatial structures of cities with less input information. Meanwhile, in general, deep learning methods usually need a large amount of computational resources and thus are more time-consuming in the training phase. Despite these limitations, by demonstrating its validity and performance, we show that the neural embedding approach offers a promising avenue for urban studies.

\section*{\sffamily Acknowledgement}
\vspace{-0.5em}
This material is based in part upon work supported by the National Science Foundation CAREER Grant CMMI-1846069, the Texas A\&M University X-Grant 699, and the Microsoft Azure AI for Public Health Grant. The authors also would like to acknowledge the data support from X-mode, Inc. and SafeGraph, Inc. Any opinions, findings, conclusions or recommendations expressed in this material are those of the authors and do not necessarily reflect the views of the National Science Foundation, Microsoft Azure, SafeGraph or X-Mode, Inc.

\section*{\sffamily Competing interests}
\vspace{-0.5em}
The authors declare that there are no competing interests.

\section*{\sffamily Data availability}
\vspace{-0.5em}
All data were collected through a CCPA and GDPR compliant framework and utilized for research purposes. The data that support the findings of this study are available from X-mode Inc., but restrictions apply to the availability of these data, which were used under license for the current study. The data can be accessed upon request submitted on x-mode.io. Other data we use in this study are all publicly available. 

\section*{\sffamily Code availability}
\vspace{-0.5em}
The code that supports the findings of this study is available from the corresponding author upon request.

\renewcommand{\refname}{\large \sffamily References}
\bibliographystyle{unsrt}  
\bibliography{references}  

\begin{thebibliography}{10}

\bibitem{Bassolas2019}
Aleix Bassolas, Hugo Barbosa-Filho, Brian Dickinson, Xerxes Dotiwalla, Paul
  Eastham, Riccardo Gallotti, Gourab Ghoshal, Bryant Gipson, Surendra~A
  Hazarie, Henry Kautz, Onur Kucuktunc, Allison Lieber, Adam Sadilek, and
  Jos{\'{e}}~J Ramasco.
\newblock {Hierarchical organization of urban mobility and its connection with
  city livability}.
\newblock {\em Nature Communications}, 10(1):4817, 2019.

\bibitem{Zhong2015}
Chen Zhong, Markus Schl{\"{a}}pfer, Stefan {M{\"{u}}ller Arisona}, Michael
  Batty, Carlo Ratti, and Gerhard Schmitt.
\newblock {Revealing centrality in the spatial structure of cities from human
  activity patterns}.
\newblock {\em Urban Studies}, 54(2):437--455, oct 2015.

\bibitem{DiClemente2018}
Riccardo {Di Clemente}, Miguel Luengo-Oroz, Matias Travizano, Sharon Xu, Bapu
  Vaitla, and Marta~C Gonz{\'{a}}lez.
\newblock {Sequences of purchases in credit card data reveal lifestyles in
  urban populations}.
\newblock {\em Nature Communications}, 9(1):3330, 2018.

\bibitem{Yang2020}
Y~Yang, C~Zhang, C~Fan, A~Mostafavi, and X~Hu.
\newblock {Towards Fairness-Aware Disaster Informatics: an Interdisciplinary
  Perspective}.
\newblock {\em IEEE Access}, 8:201040--201054, 2020.

\bibitem{Brelsford2018}
Christa Brelsford, Taylor Martin, Joe Hand, and Lu{\'{i}}s~M.A. Bettencourt.
\newblock {Toward cities without slums: Topology and the spatial evolution of
  neighborhoods}.
\newblock {\em Science Advances}, 4(8):1--9, 2018.

\bibitem{Li2017}
Ruiqi Li, Lei Dong, Jiang Zhang, Xinran Wang, Wen-Xu Wang, Zengru Di, and
  H~Eugene Stanley.
\newblock {Simple spatial scaling rules behind complex cities}.
\newblock {\em Nature Communications}, 8(1):1841, 2017.

\bibitem{Zhang2016}
Wenjia Zhang and Kara~M Kockelman.
\newblock {Congestion pricing effects on firm and household location choices in
  monocentric and polycentric cities}.
\newblock {\em Regional Science and Urban Economics}, 58:1--12, 2016.

\bibitem{Zhang2017b}
Tinglin Zhang, Bindong Sun, and Wan Li.
\newblock {The economic performance of urban structure: From the perspective of
  Polycentricity and Monocentricity}.
\newblock {\em Cities}, 68:18--24, 2017.

\bibitem{Huai2021}
Yue Huai, Hong~K Lo, and Ka~Fai Ng.
\newblock {Monocentric versus polycentric urban structure: Case study in Hong
  Kong}.
\newblock {\em Transportation Research Part A: Policy and Practice},
  151:99--118, 2021.

\bibitem{Burger2011}
Martijn Burger and Evert Meijers.
\newblock {Form Follows Function? Linking Morphological and Functional
  Polycentricity}.
\newblock {\em Urban Studies}, 49(5):1127--1149, jun 2011.

\bibitem{Midelfart2000}
K~H Midelfart, Henry Overman, Stephen Redding, and Anthony Venables.
\newblock {The location of European industry}.
\newblock Technical Report 142, apr 2000.

\bibitem{Pereira2013}
Rafael Henrique~Moraes Pereira, Vanessa Nadalin, Leonardo Monasterio, and Pedro
  H~M Albuquerque.
\newblock {Urban Centrality: A Simple Index}.
\newblock {\em Geographical Analysis}, 45(1):77--89, 2013.

\bibitem{Lu2018}
Liqun Lu, Xin Wang, Yanfeng Ouyang, Jeanne Roningen, Natalie Myers, and George
  Calfas.
\newblock {Vulnerability of Interdependent Urban Infrastructure Networks:
  Equilibrium after Failure Propagation and Cascading Impacts}.
\newblock {\em Computer-Aided Civil and Infrastructure Engineering},
  33(4):300--315, 2018.

\bibitem{Song2010a}
Chaoming Song, Tal Koren, Pu~Wang, and Albert-L{\'{a}}szl{\'{o}}
  Barab{\'{a}}si.
\newblock {Modelling the scaling properties of human mobility}.
\newblock {\em Nature Physics}, 6(10):818--823, 2010.

\bibitem{Nilsson2014}
Pia Nilsson.
\newblock {Natural amenities in urban space – A geographically weighted
  regression approach}.
\newblock {\em Landscape and Urban Planning}, 121:45--54, 2014.

\bibitem{Noulas2012}
Anastasios Noulas, Salvatore Scellato, Renaud Lambiotte, Massimiliano Pontil,
  and Cecilia Mascolo.
\newblock {A tale of many cities: Universal patterns in human urban mobility}.
\newblock {\em PLoS ONE}, 7(5):e37027, may 2012.

\bibitem{Ren2014}
Yihui Ren, M{\'{a}}ria Ercsey-Ravasz, Pu~Wang, Marta~C Gonz{\'{a}}lez, and
  Zolt{\'{a}}n Toroczkai.
\newblock {Predicting commuter flows in spatial networks using a radiation
  model based on temporal ranges}.
\newblock {\em Nature Communications}, 5(1):5347, 2014.

\bibitem{Zhou2007}
Xuesong Zhou and Hani~S Mahmassani.
\newblock {A structural state space model for real-time traffic
  origin–destination demand estimation and prediction in a day-to-day
  learning framework}.
\newblock {\em Transportation Research Part B: Methodological}, 41(8):823--840,
  2007.

\bibitem{PhysRevE.96.052301}
Xianyuan Zhan, Satish~V Ukkusuri, and P~Suresh~C Rao.
\newblock {Dynamics of functional failures and recovery in complex road
  networks}.
\newblock {\em Pysical Review E}, 96(5):52301, nov 2017.

\bibitem{Jia2021}
Tao Jia, Xi~Luo, and Xin Li.
\newblock {Delineating a hierarchical organization of ranked urban clusters
  using a spatial interaction network}.
\newblock {\em Computers, Environment and Urban Systems}, 87:101617, 2021.

\bibitem{Lammer2006}
Stefan L{\"{a}}mmer, Bj{\"{o}}rn Gehlsen, and Dirk Helbing.
\newblock {Scaling laws in the spatial structure of urban road networks}.
\newblock {\em Physica A: Statistical Mechanics and its Applications},
  363(1):89--95, 2006.

\bibitem{Pan2013}
Wei Pan, Gourab Ghoshal, Coco Krumme, Manuel Cebrian, and Alex Pentland.
\newblock {Urban characteristics attributable to density-driven tie formation}.
\newblock {\em Nature Communications}, 4(1):1961, 2013.

\bibitem{Yan2017a}
Xiao~Yong Yan, Wen~Xu Wang, Zi~You Gao, and Ying~Cheng Lai.
\newblock {Universal model of individual and population mobility on diverse
  spatial scales}.
\newblock {\em Nature Communications}, 8(1):1--9, 2017.

\bibitem{Li2014}
Yingcheng Li, Xingping Wang, Qiushi Zhu, and Hu~Zhao.
\newblock {Assessing the spatial and temporal differences in the impacts of
  factor allocation and urbanization on urban–rural income disparity in
  China, 2004–2010}.
\newblock {\em Habitat International}, 42:76--82, 2014.

\bibitem{Jin2020}
Mengjie Jin, Kun-Chin Lin, Wenming Shi, Paul T~W Lee, and Kevin~X Li.
\newblock {Impacts of high-speed railways on economic growth and disparity in
  China}.
\newblock {\em Transportation Research Part A: Policy and Practice},
  138:158--171, 2020.

\bibitem{Wong2021}
Zoey Wong, Rongrong Li, Yidie Zhang, Qunxi Kong, and Molly Cai.
\newblock {Financial services, spatial agglomeration, and the quality of urban
  economic growth–based on an empirical analysis of 268 cities in China}.
\newblock {\em Finance Research Letters}, page 101993, 2021.

\bibitem{Chen2021}
Xin Chen, Chao Xuan, and Rui Qiu.
\newblock {Understanding spatial spillover effects of airports on economic
  development: New evidence from China's hub airports}.
\newblock {\em Transportation Research Part A: Policy and Practice},
  143:48--60, 2021.

\bibitem{Cao2016}
Guangzhong Cao, Qiujie Shi, and Tao Liu.
\newblock {An integrated model of urban spatial structure: Insights from the
  distribution of floor area ratio in a Chinese city}.
\newblock {\em Applied Geography}, 75:116--126, 2016.

\bibitem{Maurer2002}
Brian~A Maurer and Mark~L Taper.
\newblock {Connecting geographical distributions with population processes}.
\newblock {\em Ecology Letters}, 5(2):223--231, mar 2002.

\bibitem{Yan2020}
Siqi Yan, Jianchao Peng, and Qun Wu.
\newblock {Exploring the non-linear effects of city size on urban industrial
  land use efficiency: A spatial econometric analysis of cities in eastern
  China}.
\newblock {\em Land Use Policy}, 99:104944, 2020.

\bibitem{Lin2016a}
Tao Lin, Caige Sun, Xinhu Li, Qianjun Zhao, Guoqin Zhang, Rubing Ge, Hong Ye,
  Ning Huang, and Kai Yin.
\newblock {Spatial pattern of urban functional landscapes along an
  urban–rural gradient: A case study in Xiamen City, China}.
\newblock {\em International Journal of Applied Earth Observation and
  Geoinformation}, 46:22--30, 2016.

\bibitem{Liu2018vertical}
Crocker~H Liu, Stuart~S Rosenthal, and William~C Strange.
\newblock {The vertical city: Rent gradients, spatial structure, and
  agglomeration economies}.
\newblock {\em Journal of Urban Economics}, 106:101--122, 2018.

\bibitem{Li2019}
Han Li, Yehua~Dennis Wei, Yangyi Wu, and Guang Tian.
\newblock {Analyzing housing prices in Shanghai with open data: Amenity,
  accessibility and urban structure}.
\newblock {\em Cities}, 91:165--179, 2019.

\bibitem{Fan2021fine}
Chao Fan, Ronald Lee, Yang Yang, and Ali Mostafavi.
\newblock {Fine-grained data reveal segregated mobility networks and
  opportunities for local containment of COVID-19}.
\newblock {\em Scientific Reports}, 11(1):16895, 2021.

\bibitem{Jia2020}
Jayson~S Jia, Xin Lu, Yun Yuan, Ge~Xu, Jianmin Jia, and Nicholas~A Christakis.
\newblock {Population flow drives spatio-temporal distribution of COVID-19 in
  China}.
\newblock {\em Nature}, 2020.

\bibitem{Li2021disparate}
Qingchun Li, Liam Bessell, Xin Xiao, Chao Fan, Xinyu Gao, and Ali Mostafavi.
\newblock {Disparate patterns of movements and visits to points of interest
  located in urban hotspots across US metropolitan cities during COVID-19}.
\newblock {\em Royal Society Open Science}, 8(1):201209, 2021.

\bibitem{Fan2021colocation}
Chao Fan, Sanghyeon Lee, Yang Yang, Bora Oztekin, Qingchun Li, and Ali
  Mostafavi.
\newblock {Effects of population co-location reduction on cross-county
  transmission risk of COVID-19 in the United States}.
\newblock {\em Applied Network Science}, 6(1):14, 2021.

\bibitem{Peng2021}
Hao Peng, Qing Ke, Ceren Budak, Daniel~M Romero, and Yong-Yeol Ahn.
\newblock {Neural embeddings of scholarly periodicals reveal complex
  disciplinary organizations}.
\newblock {\em Science Advances}, 7(17):eabb9004, apr 2021.

\bibitem{levy2014neural}
Omer Levy and Yoav Goldberg.
\newblock Neural word embedding as implicit matrix factorization.
\newblock {\em Advances in neural information processing systems},
  27:2177--2185, 2014.

\bibitem{lau-baldwin-2016-empirical}
Jey~Han Lau and Timothy Baldwin.
\newblock {An Empirical Evaluation of doc2vec with Practical Insights into
  Document Embedding Generation}.
\newblock In {\em Proceedings of the 1st Workshop on Representation Learning
  for {\{}NLP{\}}}, pages 78--86, Berlin, Germany, aug 2016. Association for
  Computational Linguistics.

\bibitem{Barkan2016}
O~Barkan and N~Koenigstein.
\newblock {ITEM2VEC: Neural item embedding for collaborative filtering}.
\newblock In {\em 2016 IEEE 26th International Workshop on Machine Learning for
  Signal Processing (MLSP)}, pages 1--6, 2016.

\bibitem{Scarselli2009}
F~Scarselli, M~Gori, A~C Tsoi, M~Hagenbuchner, and G~Monfardini.
\newblock {The Graph Neural Network Model}.
\newblock {\em IEEE Transactions on Neural Networks}, 20(1):61--80, 2009.

\bibitem{grover2016}
Aditya Grover and Jure Leskovec.
\newblock {Node2vec: Scalable Feature Learning for Networks}.
\newblock In {\em Proceedings of the 22nd ACM SIGKDD International Conference
  on Knowledge Discovery and Data Mining}, KDD '16, pages 855--864, New York,
  NY, USA, 2016. Association for Computing Machinery.

\bibitem{Hamilton2017}
William~L. Hamilton, Rex Ying, and Jure Leskovec.
\newblock {Representation Learning on Graphs: Methods and Applications}.
\newblock pages 1--23, 2017.

\bibitem{hamilton2017representation}
William~L Hamilton, Rex Ying, and Jure Leskovec.
\newblock Representation learning on graphs: Methods and applications.
\newblock {\em arXiv preprint arXiv:1709.05584}, 2017.

\bibitem{Liu2019}
L~Liu, Z~Qiu, G~Li, Q~Wang, W~Ouyang, and L~Lin.
\newblock {Contextualized Spatial–Temporal Network for Taxi
  Origin-Destination Demand Prediction}.
\newblock {\em IEEE Transactions on Intelligent Transportation Systems},
  20(10):3875--3887, 2019.

\bibitem{Cabaneros2019}
Sheen~Mclean Cabaneros, John~Kaiser Calautit, and Ben~Richard Hughes.
\newblock {A review of artificial neural network models for ambient air
  pollution prediction}.
\newblock {\em Environmental Modelling {\&} Software}, 119:285--304, 2019.

\bibitem{Peng2020}
Hao Peng, Hongfei Wang, Bowen Du, Md~Zakirul~Alam Bhuiyan, Hongyuan Ma, Jianwei
  Liu, Lihong Wang, Zeyu Yang, Linfeng Du, Senzhang Wang, and Philip~S Yu.
\newblock {Spatial temporal incidence dynamic graph neural networks for traffic
  flow forecasting}.
\newblock {\em Information Sciences}, 521:277--290, 2020.

\bibitem{yuan2021spatiotemporal}
Faxi Yuan, Yuanchang Xu, Qingchun Li, and Ali Mostafavi.
\newblock {Spatio-Temporal Graph Convolutional Networks for Road Network
  Inundation Status Prediction during Urban Flooding}, 2021.

\bibitem{YanChan2013}
Kit {Yan Chan} and Le~Jian.
\newblock {Identification of significant factors for air pollution levels using
  a neural network based knowledge discovery system}.
\newblock {\em Neurocomputing}, 99:564--569, 2013.

\bibitem{Weng2017}
Q~Weng, Z~Mao, J~Lin, and W~Guo.
\newblock {Land-Use Classification via Extreme Learning Classifier Based on
  Deep Convolutional Features}.
\newblock {\em IEEE Geoscience and Remote Sensing Letters}, 14(5):704--708,
  2017.

\bibitem{Wang_Li_Rajagopal_2020}
Zhecheng Wang, Haoyuan Li, and Ram Rajagopal.
\newblock {Urban2Vec: Incorporating Street View Imagery and POIs for
  Multi-Modal Urban Neighborhood Embedding}.
\newblock {\em Proceedings of the AAAI Conference on Artificial Intelligence},
  34(01):1013--1020, 2020.

\bibitem{Hu2021}
Sheng Hu, Song Gao, Liang Wu, Yongyang Xu, Ziwei Zhang, Haifu Cui, and Xi~Gong.
\newblock {Urban function classification at road segment level using taxi
  trajectory data: A graph convolutional neural network approach}.
\newblock {\em Computers, Environment and Urban Systems}, 87:101619, 2021.

\bibitem{x-mode}
X-Mode.
\newblock {X-Mode | Empowering Innovation with Quality Location.}, 2013.

\bibitem{Moro2021}
Esteban Moro, Dan Calacci, Xiaowen Dong, and Alex Pentland.
\newblock {Mobility patterns are associated with experienced income segregation
  in large US cities}.
\newblock {\em Nature Communications}, 12(1):4633, 2021.

\bibitem{UnitedStatesCensusBureau2019}
{United States Census Bureau}.
\newblock {American Community Survey 2014-2018 5-Year Estimates}, dec 2019.

\bibitem{SafeGraph2020}
SafeGraph.
\newblock {SafeGraph: Places Data {\&} Foot-Traffic Insights}, 2020.

\bibitem{naics}
United States~Census Bureau.
\newblock {North American Industry Classification System (NAICS).}, 2017.

\bibitem{schlichtkrull2018modeling}
Michael Schlichtkrull, Thomas~N Kipf, Peter Bloem, Rianne Van Den~Berg, Ivan
  Titov, and Max Welling.
\newblock Modeling relational data with graph convolutional networks.
\newblock In {\em European semantic web conference}, pages 593--607. Springer,
  2018.

\bibitem{Yang2015EmbeddingEA}
B~Yang, Wen-tau Yih, X~He, Jianfeng Gao, and L~Deng.
\newblock {Embedding Entities and Relations for Learning and Inference in
  Knowledge Bases}.
\newblock {\em CoRR}, abs/1412.6, 2015.

\bibitem{Murtagh2014}
Fionn Murtagh and Pierre Legendre.
\newblock {Ward's Hierarchical Agglomerative Clustering Method: Which
  Algorithms Implement Ward's Criterion?}
\newblock {\em Journal of Classification}, 31(3):274--295, 2014.

\bibitem{Ward1963}
Joe~H Ward.
\newblock {Hierarchical Grouping to Optimize an Objective Function}.
\newblock {\em Journal of the American Statistical Association},
  58(301):236--244, mar 1963.

\bibitem{Xue2021QuantifyingSH}
Jiawei Xue, Nan Jiang, Senwei Liang, Qi~Pang, S~Ukkusuri, and Jianzhu Ma.
\newblock {Quantifying spatial homogeneity of urban road networks via graph
  neural networks}.
\newblock {\em ArXiv}, abs/2101.0, 2021.

\bibitem{McInnes2018UMAPUM}
L~McInnes and John Healy.
\newblock {UMAP: Uniform Manifold Approximation and Projection for Dimension
  Reduction}.
\newblock {\em ArXiv}, abs/1802.0, 2018.

\bibitem{Wang2021}
Juexin Wang, Anjun Ma, Yuzhou Chang, Jianting Gong, Yuexu Jiang, Ren Qi, Cankun
  Wang, Hongjun Fu, Qin Ma, and Dong Xu.
\newblock {scGNN is a novel graph neural network framework for single-cell
  RNA-Seq analyses}.
\newblock {\em Nature Communications}, 12(1):1882, 2021.

\end{thebibliography}







\end{document}